\documentclass[12pt]{article}

\usepackage{amssymb,amsmath}

\begin{document}

\begin{titlepage}

\begin{flushright}
arXiv:2309.05648
\end{flushright}
\vskip 2.5cm

\begin{center}
{\Large \bf Radiation from an Oscillating Dipole in the Presence\\
of Photon-Sector CPT and Lorentz Violation}
\end{center}

\vspace{1ex}

\begin{center}
{\large Joshua O'Connor and Brett Altschul\footnote{{\tt altschul@mailbox.sc.edu}}}

\vspace{5mm}
{\sl Department of Physics and Astronomy} \\
{\sl University of South Carolina} \\
{\sl Columbia, SC 29208} \\
\end{center}

\vspace{2.5ex}

\medskip

\centerline {\bf Abstract}

\bigskip

We examine one of the standard loci for studying electromagnetic wave emission---the
radiation from an oscillating electric dipole---in a model in which the electromagnetic
sector is modified to include novel CPT- and Lorentz-violating propagation effects
involving a preferred axial vector background. We evaluate
the vacuum-birefringent radiation fields, including nonperturbative terms where appropriate.
In general, the energy-momentum carried by the fields in this model is known to have a complicated
nonperturbative structure, which cannot be captured by naive power series expansions
in the components of the preferred background vector.
However, we nevertheless find that at the lowest nontrivial orders, there are actually no
modifications to the Larmor expressions for the energy-momentum emission.

\bigskip

\end{titlepage}

\newpage

\section{Introduction}

The special theory of relativity underlies all of fundamental physics as we currently understand
it---including the general theory, as well as relativistic quantum field theory. Moreover, the development
of relativity contributed tremendously to our understanding of the critical importance of symmetries in
physics. Nevertheless, there have always been questions---both theoretical and empirical---about whether
special relativity as it was introduced by Einstein in 1905 truly represents an exact local symmetry
structure for spacetime, or whether it is merely an extremely accurate approximate model.
In the twentieth century and beyond, the study of apparent symmetries that
are eventually discovered to be not exactly but only approximately valid has become extremely important and
has provided many fruitful insights about the fundamental interactions of nature.

However, experimental tests of relativity and theoretical analyses of test theories with broken
Lorentz symmetry were not really approached in a systematic fashion until the 1990s. Using modern
effective field theory (EFT), it became comparatively straightforward to parameterize very
general test theories for Lorentz violation in particle physics and gravitation. It turns out that
these systematic EFTs allow for much wider arrays of types of anisotropy and Lorentz boost violations
than had previously been examined. These theoretical developments were followed by an upsurge in
experimental interest in Lorentz symmetry tests, because it was realized that there was a much broader
landscape of potentially symmetry-violating phenomena. So far, the new generations of experiments have
not found any convincing evidence of Lorentz violation, but the increasingly precise tests have
continued to be a significant area of research. One key reason for the continued interest is
that, however unlikely Lorentz violation is deemed to be, if it is ever confirmed experimentally,
that would be such a profound discovery that it would change a lot of what we think we understand about
the fundamental nature of the universe we live in.

It is now well understood how to set up the general local EFT that describes
Lorentz-violating modifications to the physics of known standard model species~\cite{ref-kost1,ref-kost2}.
This EFT, known as the standard model extension (SME), is also capable of describing all
stable, unitary, and local forms of CPT violation, because of the close connections between CPT violation
and Lorentz violation in theories with well-defined $S$-matrices~\cite{ref-greenberg}.
Moreover, the SME can also be expanded to cover gravitation, although the extension to metric theories of
gravity creates additional complications beyond those seen in the particle physics sector. Part of the
reason for this is that the particle sector of the SME is formulated using the
language of quantum field theory (QFT), just like the standard model, while metric theories of
gravity (like general relativity and its generalizations) are not really understood beyond the
classical level.

In many situations (such as interpreting the results of laboratory Lorentz and CPT tests), it makes sense
to consider only a truncated version of the SME. The most important such truncation is the
minimal SME. The minimal SME is the subsector of the theory that is expected to be
renormalizable, because it contains only a finite number of local, gauge-invariant operators which are
constructed from standard-model scalar, spinor, and vector fields, and which have dimensions of
$($momentum$)^{4}$ or less. These operators resemble those in the conventional standard model
Lagrange density, except that they possess additional Lorentz indices that are not contracted with
other dynamical quantities but with preferred background vectors and tensors. When dealing with
electromagnetic phenomena, it is typical to truncate the minimal SME even further, to just a minimal
Lorentz- and CPT-violating extension of quantum electrodynamics (QED).

Although the minimal SME and the minimal QED extension are quantum theories, they
exhibit many potentially novel phenomena already at the classical level. In particular, radiation
emission may be heavily modified by the presence of the symmetry-breaking terms in the action. For example,
in theories in which the maximum speeds of all species are not equal, it is easy to envision that there
could be Cerenkov radiation in vacuum.
This
paper will look at another radiation process that is particularly simple in the Maxwell theory---emission
by a harmonically oscillating dipole.

While a completely new effect that can only occur because of Lorentz violation (such
as vacuum Cerenkov radiation when boost symmetry is broken, or a transition between two states with different
angular momentum values in a theory with broken rotation symmetry) will typically appear at
second order in the symmetry-breaking interaction coefficients, modifications to phenomena
that already occur in the standard theory can be observable at lower order---for example,
as small interference effects on top of conventional observables. Our approach in this paper will
be to look for a modification of this nature---a change to the conventional Larmor expression
for the energy-momentum radiated by a harmonically oscillating dipole. Radiation spectra are
known to be substantially and nonperturbatively modified at second order in the magnitude of
a Lorentz-violating Chern-Simons term in the photon sector. However, by looking for modifications
to standard dipole radiation, we open up the possibility of finding observable changes already
at first order. Dipole radiation also provides an extremely clean theoretical laboratory for identifying
novel behavior. For example, radiation damping is typically a very awkward topic in classical
electrodynamics; however, for the dipole source created by a harmonically oscillating charge, it
is possible to define a radiative friction term which (so long as the radiation is not too
rapid) avoids most of the awkwardness that typically accompanies the evaluation of the self-force
on an accelerated charge.

This paper is organized as follows.  In section~\ref{sec-model}, we describe the model of
Lorentz-violating electrodynamics with a CPT-odd Chern-Simons term. In this theory,
the free propagation models of the electromagnetic field exhibit vacuum birefringence. Taking what is
known about these plane wave modes, we determine the lowest-order modifications to the radiation-zone fields
of an oscillating dipole in section~\ref{sec-fields} and evaluate the standard Poynting vector
$\vec{S}^{(0)}=\vec{E}\times\vec{B}$ with the modified fields. Then, in section~\ref{sec-further}, we look
at two additional ways in which the energy and momentum emission may be modified, which were not captured
by the first set of calculations. Finally, section~\ref{sec-concl} presents our conclusions and the
outlook for further extensions of this work.

\section{CPT- and Lorentz-Violating Electrodynamics}

\label{sec-model}

In the minimal SME, the Lagrange density for the electromagnetic sector is~\cite{ref-kost1,ref-kost2}
\begin{equation}
\mathcal{L}=-\frac{1}{4}F^{\mu\nu}F_{\mu\nu}
-\frac{1}{4}k_{F}^{\mu\nu\rho\sigma}F_{\mu\nu}F_{\rho\sigma}
+\frac{1}{2}k_{AF}^{\mu}\epsilon_{\mu\nu\rho\sigma}F^{\nu\rho}A^{\sigma}
-j^{\mu}A_{\mu}.
\end{equation}
This includes all the superficially renormalizable
operators that can be constructed solely out of photon fields. The CPT-even operators
are the ones that multiply the nineteen independent $k_{F}^{\mu\nu\rho\sigma}$
coefficients. Although there are many potentially interesting phenomena that could appear
in the presence of nonzero $k_{F}^{\mu\nu\rho\sigma}$ terms, we shall not be focusing on them here.
Instead we shall be looking at possible effects of the four-component (axial vector)
$k_{AF}^{\mu}$ term.  The associated operators are all CPT odd, and $k_{AF}^{\mu}$ itself
has dimension $($momentum$)^{1}$.

Up-to-date bounds on the SME coefficients may be found in Ref.~\cite{ref-tables}.
The terms in the minimal SME action that produce vacuum birefringence have been tightly bounded
using polarimetric data from cosmologically-distant sources. For the four
components of $k^{\mu}_{AF}$, which give rise to wavelength-independent rotations in the planes of
polarization of initially linear-polarized waves and parity-violating
correlations in the polarization of the cosmic microwave background (CMB), the bounds
have been placed at the $10^{-44}$ GeV level or better,
beginning with Ref.~\cite{ref-carroll1} for quasar jets and more recently using CMB data~\cite{ref-caloni}.
Of the nineteen independent coefficients in $k_{F}^{\mu\nu\rho\sigma}$,
ten of them also generate photon birefringence, and they are also quite tightly constrained, at the
$10^{-34}$--$10^{-38}$~\cite{ref-exirifard,ref-mewes8,ref-gerasimov}
levels. On the other hand, the remaining nine coefficients from
$k_{F}^{\mu\nu\rho\sigma}$ are much more difficult to measure, and precision optical experiments
and astrophysical data
have been used to bound them only at $10^{14}$--$10^{-22}$ levels~\cite{ref-michimura,ref-mewes9,ref-duenkel}.
However, this last group of
parameters, the nonbirefringent ones that are the most challenging to measure
directly, actually have effects on dipole
radiation that are already completely understood, since they may actually be eliminated from the photon sector
entirely by means of an oblique linear transformation of the spacetime coordinates.

In contrast, although they have already been extremely well constrained by polarimetry, the CPT-odd
$k_{AF}^{\mu}$ terms are still of interest for theoretical, but also potentially practical, reasons.
From its first introduction, there have been questions about whether it is even possible to have a nonzero
$k_{AF}^{\mu}$ in a consistent field theory, and so far there have been conflicting
indications~\cite{ref-carroll1,ref-kost20,ref-schober1,ref-karki3}.
The theory with just the Chern-Simons term does not appear to be energetically stable, even
classically, and there were
concerns about whether it was even possible for such a theory to have a well-defined, unitary
$S$-matrix---which is a fairly basic requirement for a physically meaningful
theory. One obvious way that such an inconsistency
might manifest itself would be through runaway vacuum Cerenkov radiation, because the Chern-Simons
theory contains arbitrarily slow phase speeds. In fact, any radiation process might
potentially be subject to unstable, nonperturbative behavior that might invalidate the theory, so
studying standard radiation scenarios in the presence of the Chern-Simons term could potentially lead us to
new physical conclusions about the phenomenalistic viability of the theory. This is one of the key
motivations for this work.
Moreover, questions about whether the structure of the Chern-Simons term could make a $k_{AF}^{\mu}$ theory
mathematically or physically inconsistent are actually fairly reasonable, in light of its unusual structural
properties. The term's structure means that potential radiative corrections to the photon $k_{AF}^{\mu}$
must come from virtual processes that are extremely similar to those that appear in chiral anomaly triangle
diagrams~\cite{ref-jackiw1,ref-chung1,ref-victoria2}, and the cancellation of
the related gauge anomalies
is already known to give nontrivial restrictions on the structure of internally consistent quantum
field theories.

For brevity, we shall drop the ``$AF$'' labels and henceforth write $k_{AF}^{\mu}=k^{\mu}$.
With the Chern-Simons term present, the purely electromagnetic part of the energy-momentum tensor
becomes~\cite{ref-carroll1}
\begin{equation}
\Theta^{\mu\nu}= -F^{\mu\alpha}F^{\nu}\,_{\alpha}+\frac{1}{4}g^{\mu\nu}
F^{\alpha\beta}F_{\alpha\beta} -\frac{1}{2}k^{\nu}\epsilon^{\mu\alpha\beta\gamma}
F_{\beta\gamma}A_{\alpha}.
\label{eq-Theta}
\end{equation}
The tensor is not symmetric, and the asymmetry is in fact a measure of the Lorentz violation.
The main terms of interest are the energy density
($\mathcal{E}=\Theta^{00}$), energy flux ($S_{j}=\Theta^{j0}$), and
momentum density ($\mathcal{P}_{j}=\Theta^{0j}$),
\begin{eqnarray}
\mathcal{E} & = & \frac{1}{2}\vec{E}^{2}+\frac{1}{2}\vec{B}^{2}-
k_{0}\vec{A}\cdot\vec{B}\equiv\mathcal{E}^{(0)}-k_{0}\vec{A}\cdot\vec{B}
\label{eq-E} \\
\vec{S} & = & \vec{E}\times\vec{B}-k_{0}A_{0}\vec{B}+k_{0}\vec{A}\times\vec{E}
\equiv\vec{S}^{(0)}-k_{0}A_{0}\vec{B}+k_{0}\vec{A}\times\vec{E}
\label{eq-S} \\
\vec{{\cal P}} & = & \vec{E}\times\vec{B}-\vec{k}\left(\vec{A}\cdot\vec{B}\right)
\equiv\vec{S}^{(0)}-\vec{k}\left(\vec{A}\cdot\vec{B}\right),
\label{eq-P}
\end{eqnarray}
in terms of the time and space components of $k^{\mu}=(k_{0},\vec{k}\,)$.
Note that none of these quantities are gauge invariant, because they depend not just
on the field strengths $\vec{E}$ and $\vec{B}$, but also on the scalar and vector
potentials.
However, the total energy and total momentum, found by integrating $\mathcal{E}$ and
$\vec{\mathcal{P}}$ over all space, are gauge invariant.
This is not necessarily obvious from the forms of these densities,
but the key property is that $\mathcal{E}$ and $\vec{\mathcal{P}}$ (and also the Lagrange
density $\mathcal{L}$) change under gauge transformations by terms that are total
derivatives, and which thus make no contributions to the integrated quantities.

The form of the energy density exhibits one of the elements that makes the analysis of this
theory somewhat tricky---that the energy is not bounded below. The
$-k_{0}\vec{A}\cdot\vec{B}$ term may be made arbitrarily
negative by increasing the magnitude of the field $\vec{A}$ (and thus simultaneously increasing
the magnitude of $\vec{B}=\vec{\nabla}\times\vec{A}$). For modes of the field with sufficiently
long wavelengths, the $-k_{0}\vec{A}\cdot\vec{B}$ in $\mathcal{E}$ will win out over the usual
$\frac{1}{2}\vec{B}^{2}$.
This downward unboundedness of the the energy also has a manifestation in the plane wave dispersion
relations.
If $k^{\mu}$ is purely spacelike, meaning $\vec{k}=0$, the (birefringent) dispersion relation
is $\omega_{\pm}^{2}=Q(Q\pm2k_{0})$, for waves with wave vector $\vec{Q}$ and positive and
negative helicities, respectively. This is a special case of (\ref{eq-Q-exact}) below,
and it identifies the Fourier modes with $Q<2|k_{0}|$ as
precisely those for which instabilities may occur. For these modes, the time dependence may
be exponentially growing, rather than oscillatory. This kind of 
runaway growth is ``powered'' by the
ever-decreasing energy of the $-k_{0}\vec{A}\cdot\vec{B}$ term in $\mathcal{E}$.

These runaway modes do not necessarily have to ruin the Chern-Simons theory with timelike
$k^{\mu}$, but they certainly do raise significant questions. It is possible to avoid the
instability by giving up causality. The Green's functions for the theory may be chosen that the
exponentially growing modes are never populated, but the cost is having solutions in which
charged sources will begin to radiate before they actually begin to move~\cite{ref-carroll1}.
The acausality is
relatively weak so long as $k_{0}$ is small, so that using the acausal Green's functions
to describe the emission of long radio-frequency wave trains is probably unproblematic.
However, it remains unclear whether, at a more fundamental level,
it is physically sensible to have a theory with these kinds of acausal behavior.
Perhaps even more strangely, the modes with $Q<2|k_{0}|$ actually seem to rescue the timelike
Chern-Simons theory from other kinds of instability, rather than causing it! Any
photon dispersion relation for with $\omega/Q<1$ opens up the possibility of Cerenkov radiation.
Conventional Cerenkov radiation is a phenomenon that occurs in material media, in which
the phase speed of light is slowed down. When charged particles move through faster than
the in-medium $\omega/Q$, they emit a burst of radiation, analogous to the masses moving
faster than the speed of sound in a fluid produce sonic booms.
Since the real branch of the Chern-Simons dispersion relation $\omega^{2}=Q(Q-2|k_{0}|)$
extends all the way down to $\omega=0$, any moving charge with speed $v$ is going to outpace some of
the propagating electromagnetic field modes, no matter how small $v$ is.
A natural expectation (based on phase space
availability considerations~\cite{ref-altschul12}) would therefore be that any
charge in uniform motion would lose energy to vacuum Cerenkov radiation, until it came to rest
in the frame where $\vec{k}=0$. However, this turns out not to be case, for the
following subtle reason~\cite{ref-schober1,ref-altschul21,ref-decosta1}.
Modes of the field with $2|k_{0}|<Q<2|k_{0}|/(1-v^{2})$ do carry power away, but the modes
with $Q<2|k_{0}|$ actually carry away negative power, in an amount that exactly cancels the net
positive emission. For the fields following a charged particle in uniform motion,
the $\omega^{2}<0$ modes are actually associated with propagating solutions carrying negative
energies. The result is that
in the timelike Chern-Simons theory, just as in standard electrodynamics, charges in uniform motion
lose no net energy via radiation.

The theory with a spacelike Chern-Simons four-vector $k^{\mu}$ is potentially better behaved
than the timelike theory, but it has other, closely related, peculiarities. The energy instability,
which is unavoidable if $k^{\mu}$ is timelike, instead depends on the choice of frame. In particular,
there is only an obvious vacuum state, with the energy density function $\mathcal{E}$ bounded from below,
in the $k_{0}$ reference frame. Analyzing the theory in this preferred frame (including potentially
quantizing it) looks like it may be relatively unproblematic, but this leaves open the question of how
the theory should be quantized in a different frame, in which the
$-k_{0}\vec{A}\cdot\vec{B}$ term in $\mathcal{E}$ is present. The existence of the particular frame in
which the the theory is manifestly stable energetically actually has important implications for the vacuum
Cerenkov radiation in the spacelike theory. Once again, a charged particle with a constant velocity
$\vec{v}$ may emit radiation. This radiation tends to damp out the motion, until the charge is at rest in
the frame where $k^{\mu}$ has no temporal component~\cite{ref-lehnert2}. This is a remarkable explicit
connection between the energetic stability condition and the dynamics of individual charges, and
this connection was a very important motivation for the current work.

Understanding radiation processes in the Chern-Simons theory is tricky, and there have
already been a number of different innovative approaches. For the timelike
theory, the original derivation of the
acausal Green's functions was a major step, since it showed that there were systems that obeyed
the equations of motion and emitted radiation, without the amplitudes of
long-wavelength modes growing uncontrollably. However, the acausality was an obvious drawback.
Looking specifically at
the vacuum Cerenkov radiation emitted by a charge in uniform motion provided one way out
of this dilemma. If the electromagnetic field, including the radiation component, is moving along
in synchronization with the source charge, the time dependences of the field components can be
inferred from their spatial profiles at a fixed time, leaving no room of runaway behavior.
Considering only versions of the theory with spacelike $k^{\mu}$ also eliminates the obvious stability
problem, but only in one frame, and the radiation rate calculated in Ref.~\cite{ref-lehnert2}
takes an extremely unusual and manifestly nonperturbative form---meaning that the effect cannot
be found by expanding the theory in powers of the space and time components of the
background four-vector $k^{\mu}$. Instead, the rate of momentum emission via the
vacuum Cerenkov radiation coming from a stationary charge is proportional to
$-k_{0}\vec{k}|k_{0}|^{3}/|\vec{k}|^{3}$. (However, in contrast, in the presence of the other
types of Lorentz violation, vacuum Cerenkov radiation is more typically a threshold effect, as it is in
matter~\cite{ref-altschul9,ref-anselmi})

When investigating a modified version of a generally well-understood theory, looking at
modifications to effects that are permitted in the standard theory can, in many cases, be more
fruitful than studying entirely new phenomena. Vacuum Cerenkov radiation would be a strikingly
novel phenomenon if it were ever to be observed, but the emission rate may be very small.
In the spacelike $k^{\mu}$ theory, the
spontaneous force---something which does not exist at all in conventional
electrodynamics---is of $\mathcal{O}\left(k_{0}^{4}/\vec{k}^{2}\right)$. This work was, in part,
an outgrowth of the hope that by looking at a different radiation process---one which does exist in
standard electrodynamics---we might uncover qualitatively similar
radiation effects, but that the mathematical
descriptions of the modified radiation could be found by expanding all quantities to leading
order in $k^{\mu}$.

We shall therefore consider the radiation emitted by a very simple system---a harmonically oscillating
electric dipole
$\vec{p}$. There are several different ways of expressing the charge and current densities of a
radiating source in terms of sums (or integrals) of simple basis functions. Thanks to the linearity of
electrodynamics, the resulting electromagnetic fields will be coherent sums of the fields produced
by just the basis functions. For example, the
radiation fields generated by an accelerated charge may be written as integrals of the
radiation fields (with kinked electric and magnetic field lines)
produced by instantaneous impulsive accelerations. However, this method would probably be inapt in the
general Chern-Simons theory theory, because it provides no evident way to control the excitation
of unstable long-wavelength modes. In contrast, a description of the sources and fields using a
Fourier transform in the time domain deals with the problem quite neatly. By working with a source
that is excited solely at a frequency $\omega$, we can be assured that no modes with other time
dependences will be excited. Moreover, the presence of $\omega$ serves another mathematical purpose.
In the analyses of vacuum Cerenkov radiation, the only physical quantities on which the radiation
rate can depend are the charge $q$, its velocity $\vec{v}$ (both dimensionless), and the
Lorentz-violating $k^{\mu}$, which has units of $($momentum$)^{1}$. Since the power emitted has
units of $($momentum$)^{2}$, it must---for dimensional reasons alone---be a homogeneous function of
degree two of the components of $k^{\mu}$. It is simply not possible to have an effect that appears at
perturbative $\mathcal{O}(k)$. However, with the oscillating dipole, there are additional dimensional
quantities involved. Multipole moments have units of $($momentum$)^{-\ell}$, but $\omega$ has units
of positive $($momentum$)^{1}$. That makes it possible to have radiation emission effects at only
linear order in $k^{\mu}$.
The standard radiation fields may mix with the $k^{\mu}$-modified fields to produce correction
terms of $\mathcal{O}(k\omega^{3})$.

So for our source, we shall take an oscillating electric dipole at the origin and extrapolate its
fields outwards into the radiation zone.
At very short distances, the sources on the right-hand side of the modified Gauss's law
\begin{equation}
\vec{\nabla}\cdot\vec{E}=\rho+2\vec{k}\cdot\vec{B}
\label{eq-Gauss}
\end{equation}
are dominated by the conventional term, since the charge density $\rho$ has a strong singularity,
$\rho=-\vec{p}\cdot\vec{\nabla}\delta^{3}(\vec{r}\,)$; and similarly for the
modified Ampere-Maxwell law,
\begin{equation}
\vec{\nabla}\times\vec{B}-\frac{\partial\vec{E}}{\partial t}=\vec{J}+2k_{0}\vec{B}
-2\vec{k}\times\vec{E},
\end{equation}
where
the singular current distribution of the oscillating dipole dominates the right-hand side in
the immediate vicinity of the dipole.

A key consequence of this is that we may use the standard forms for the oscillating
fields at infinitesimally short distances from the origin, then propagate them out to larger radii
using the known propagation characteristics of the Chern-Simons theory.  This takes advantage of the
fact that the Fourier spectrum of the propagating modes is well known. Meanwhile we shall, whenever it is
convenient, neglect any near-field terms that fall off too rapidly with distance to contribute to
the energy and momentum flows at large distances. We shall also take $k^{\mu}$ to be small, so that we
never need consider terms beyond first order in the components of $k$---except when they are multiplied
by a large distance $r$. This last caveat is necessary because we are interested in the behavior of the
fields at arbitrarily large distances. Note also that additional care may be
specifically needed for waves propagating
in a direction $\hat{r}$ for which $k^{\mu}\hat{r}_{\mu}\equiv k_{0}-\vec{k}\cdot\hat{r}\approx 0$, since
the normal mode polarization vectors may be significantly modified in these angular regions.

\section{Radiation Fields Modified by $k^{\mu}$}

\label{sec-fields}

The standard far-field form for the magnetic field of the oscillating dipole with complexified
amplitude $\vec{p}\,(t)=\vec{p}\,e^{-i\omega t}$ is
\begin{equation}
\vec{B}=\frac{\omega^{2}}{4\pi}\left(\hat{r}\times\vec{p}\,\right)\frac{e^{i\omega(r-t)}}{r}.
\end{equation}
To find the version of this in the Chern-Simons theory, it suffices to calculate the projections of
this field along the polarization eigenvectors of the modified theory and to attach to each
projection a modified propagation factor $e^{i(Qr-\omega t)}$.
When the Chern-Simons vector is purely timelike, $k^{\mu}=(k_{0},\vec{0}\,)$, the
polarization vectors for the normal modes of propagation are circular,
\begin{equation}
\hat{\epsilon}_{\pm}=\frac{1}{\sqrt{2}}\left(\hat{\theta}\pm i\hat{\phi}\right)\!.
\label{eq-epsilon-pm}
\end{equation}
When $k^{\mu}$ has spatial components as well, these are still very close to the exact polarization
vectors, differing only meaningfully around $k^{\mu}\hat{r}_{\mu}\approx 0$.
To leading order, the wave numbers corresponding to these circular polarization modes are
\begin{equation}
Q_{\pm}=\omega\pm k^{\mu}\hat{r}_{\mu}=\omega\pm\left(k_{0}-\vec{k}\cdot\hat{r}\right)\!.
\label{eq-Q-pm}
\end{equation}
This is the leading approximation to the exact (but implicit) relationship between frequency
and wave vector~\cite{ref-carroll1}
\begin{equation}
\omega^{2}-Q_{\pm}^{2}=\mp 2\left[k_{0}Q_{\pm}-\left(\vec{k}\cdot\hat{r}\right)\omega\right]
\left(1-\frac{4\left|\vec{k}\times\hat{Q}\right|^{2}}{\omega^{2}-Q_{\pm}^{2}}\right)^{-1/2}.
\label{eq-Q-exact}
\end{equation}
With the leading order $Q_{\pm}$, the propagating part of the magnetic field in the Lorentz-violating theory must be
\begin{equation}
\vec{B}=\frac{\omega^{2}}{4\pi}\frac{e^{i\omega(r-t)}}{r}
\sum_{\pm}\left[\hat{\epsilon}_{\pm}^{*}\!\cdot\left(\hat{r}\times\vec{p}\,\right)\right]
e^{\pm i\left(k_{0}r-\vec{k}\cdot\vec{r}\right)}\,\hat{\epsilon}_{\pm}
\label{eq-B-sep}
\end{equation}
(with the sum referring to the sum over the $\pm$ modes).

We shall select a coordinate system so that the oscillating dipole lies in the $xy$-plane,
\begin{equation}
\vec{p}=p_{x}\,\hat{x}+p_{y}\,\hat{y}=p_{1}\,\hat{x}+e^{i\alpha}p_{2}\,\hat{y},
\end{equation}
where $p_{1}$, $p_{2}$, and $\alpha$ are real quantities.
Although the components $p_{x}$ and $p_{y}$ would generally be complex, we have taken
advantage of the fact that we can shift the overall phase (either by a redefinition of
the zero of $t$ or by a rotation of the $xy$-plane) to make the quadrature component
along the $x$-direction real. It will also be convenient to have a separation of the
oscillating dipole moment into its phase components---that is, its real and imaginary parts,
\begin{equation}
\vec{p}=\vec{p}_{R}+i\vec{p}_{I}.
\label{eq-p-planar}
\end{equation}
Then the necessary triple products are
\begin{equation}
\hat{\epsilon}_{\pm}^{*}\!\cdot\left(\hat{r}\times\vec{p}\,\right)=
\frac{1}{\sqrt{2}}\left(\hat{\theta}\mp i\hat{\phi}\right)\!\cdot\!
\left[\left(p_{1}\sin\phi-e^{i\alpha}p_{2}\cos\phi\right)\hat{\theta}+
\left(p_{1}\cos\phi+e^{i\alpha}p_{2}\sin\phi\right)\cos\theta\,\hat{\phi}\right]\!,
\label{eq-tri-prod}
\end{equation}
so that the expressions for the full summand terms are
\begin{eqnarray}
\left[\hat{\epsilon}_{\pm}^{*}\!\cdot\left(\hat{r}\times\vec{p}\,\right)\right]
e^{\pm i\left(k_{0}r-\vec{k}\cdot\vec{r}\right)}\,\hat{\epsilon}_{\pm} & = & 
\frac{1}{2}\left(p_{1}\sin\phi-e^{i\alpha}p_{2}\cos\phi\mp ip_{1}\cos\theta\cos\phi\right. \nonumber\\
& & \left.\mp ie^{i\alpha}p_{2}\cos\theta\sin\phi\right)
e^{\pm i\left(k_{0}r-\vec{k}\cdot\vec{r}\right)}\left(\hat{\theta}\pm i\hat{\phi}\right)\!.
\end{eqnarray}
Note that the two terms (corresponding to $\pm$ subscripts) are almost complex conjugates, except
that the $e^{i\alpha}$ factors are unchanged between the two. Thus it will be convenient to write
the expressions in the forms
\begin{eqnarray}
\left[\hat{\epsilon}_{+}^{*}\!\cdot\left(\hat{r}\times\vec{p}\,\right)\right]
e^{+ i\left(k_{0}r-\vec{k}\cdot\vec{r}\right)}\,\hat{\epsilon}_{+} & = & 
\frac{1}{2}\left(\mathcal{A}_{1}\,\hat{\theta}+i\mathcal{A}_{1}\,\hat{\phi}
+e^{i\alpha}\mathcal{A}_{2}\,\hat{\theta}+ie^{i\alpha}\mathcal{A}_{2}\,\hat{\phi}\right) \\
\left[\hat{\epsilon}_{-}^{*}\!\cdot\left(\hat{r}\times\vec{p}\,\right)\right]
e^{-i\left(k_{0}r-\vec{k}\cdot\vec{r}\right)}\,\hat{\epsilon}_{-} & = & 
\frac{1}{2}\left(\mathcal{A}_{1}^{*}\,\hat{\theta}-i\mathcal{A}_{1}^{*}\,\hat{\phi}
+e^{i\alpha}\mathcal{A}_{2}^{*}\,\hat{\theta}
-ie^{i\alpha}\mathcal{A}_{2}^{*}\,\hat{\phi}\right)\!,
\end{eqnarray}
where $\mathcal{A}_{1}$ and $\mathcal{A}_{2}$ are
\begin{eqnarray}
\mathcal{A}_{1} & = & p_{1}\left(\sin\phi-i\cos\theta\cos\phi\right)
\left[\cos\left(k_{0}r-\vec{k}\cdot\vec{r}\right)+i\sin\left(k_{0}r-\vec{k}\cdot\vec{r}\right)\right] \\
& = & p_{1}\left\{\left[\sin\phi\cos\left(k_{0}r-\vec{k}\cdot\vec{r}\right)+\cos\theta\cos\phi
\sin\left(k_{0}r-\vec{k}\cdot\vec{r}\right)\right]\right. \nonumber \\
& & -\left.i\left[\cos\theta\cos\phi\cos\left(k_{0}r-\vec{k}\cdot\vec{r}\right)
-\sin\phi\sin\left(k_{0}r-\vec{k}\cdot\vec{r}\right)\right]
\right\} \\
\mathcal{A}_{2} & = & -p_{2}\left(\cos\phi+i\cos\theta\sin\phi\right)
\left[\cos\left(k_{0}r-\vec{k}\cdot\vec{r}\right)+i\sin\left(k_{0}r-\vec{k}\cdot\vec{r}\right)\right] \\
& = & p_{2}\left\{\left[-\cos\phi\cos\left(k_{0}r-\vec{k}\cdot\vec{r}\right)+\cos\theta\sin\phi
\sin\left(k_{0}r-\vec{k}\cdot\vec{r}\right)\right]\right. \nonumber \\
& & -\left.i\left[\cos\theta\sin\phi\cos\left(k_{0}r-\vec{k}\cdot\vec{r}\right)
+\cos\phi\sin\left(k_{0}r-\vec{k}\cdot\vec{r}\right)\right]
\right\}\!.
\end{eqnarray}

For a linear dipole, $\alpha=0$, these expressions reduce to a familiar birefringent form. The
$\theta$-component of the magnetic field is proportional to
$\Re\left\{\mathcal{A}_{1}+\mathcal{A}_{2}\right\}$, or 
\begin{eqnarray}
B_{\theta} & \propto & (p_{1}\sin\phi-p_{2}\cos\phi)\cos\left(k_{0}r-\vec{k}\cdot\vec{r}\right)
\nonumber\\
& & +(p_{1}\cos\phi+p_{2}\sin\phi)\cos\theta\sin\left(k_{0}r-\vec{k}\cdot\vec{r}\right)
\label{eq-B-theta-linear-2} \\
& \propto & \cos\left\{\left(k_{0}r-\vec{k}\cdot\vec{r}\right)-\tan^{-1}\left[
\frac{(p_{1}\cos\phi+p_{2}\sin\phi)\cos\theta}{p_{1}\sin\phi-p_{2}\cos\phi}\right]\right\}\!.
\label{eq-B-theta-linear}
\end{eqnarray}
Similarly, the $\phi$-component may be seen to be proportional in the same fashion to the
the imaginary part, 
\begin{eqnarray}
B_{\phi} & \propto & (p_{1}\sin\phi-p_{2}\cos\phi)\sin\left(k_{0}r-\vec{k}\cdot\vec{r}\right)
\nonumber\\
& & -(p_{1}\cos\phi+p_{2}\sin\phi)\cos\theta\cos\left(k_{0}r-\vec{k}\cdot\vec{r}\right) \\
& \propto & \sin\left\{\left(k_{0}r-\vec{k}\cdot\vec{r}\right)-\tan^{-1}\left[
\frac{(p_{1}\cos\phi+p_{2}\sin\phi)\cos\theta}{p_{1}\sin\phi-p_{2}\cos\phi}\right]\right\}\!.
\label{eq-B-phi-linear}
\end{eqnarray}
This shows a linearly polarized field of constant amplitude $\left|\vec{B}\,\right|$,
but with a polarization direction
that is corkscrewing around the propagation direction $\hat{r}$. This effect has been used as the
basis for placing extremely tight constraints on the components of the physical $k^{\mu}$
coefficients~\cite{ref-carroll1,ref-carroll2,ref-kost11,ref-kost21}.
The phase constant in (\ref{eq-B-theta-linear}) and (\ref{eq-B-phi-linear}) is
independent of the radius. It depends only on the oscillating dipole $\vec{p}$ and the angles
$(\theta,\phi)$ that describe the radiation direction. The quantities $p_{1}\cos\phi+p_{2}\sin\phi$
and $p_{1}\sin\phi-p_{2}\cos\phi$ are the projections of $\vec{p}$ parallel and perpendicular
to the axial direction $\hat{\rho}$, and the former appears with the geometrical foreshortening
factor $\cos\theta$.
Obviously, the same holds for the amplitude of the local magnetic field oscillations,
which is proportional to the square root of
\begin{equation}
(p_{1}\cos\phi+p_{2}\sin\phi)^{2}\cos^{2}\theta+(p_{1}\sin\phi-p_{2}\cos\phi)^{2}
=\vec{p}\,^{2}-\left(\vec{p}\cdot\hat{r}\right)^{2}\!,
\end{equation}
although
for this quantity the dependence on the projections of $\vec{p}$ is a well-known feature of
the standard theory.

To the extent that the polarization vectors are accurately represented by (\ref{eq-epsilon-pm})
in the Chern-Simons theory, it continues to be the case that the plane wave propagating modes
of the theory have electric and magnetic components related by $\vec{E}=\vec{B}\times\hat{r}$, so that
$E_{\theta}=B_{\phi}$ and $E_{\phi}=-B_{\theta}$.
The effect of $k^{\mu}$ is to cause the electric field direction to revolve in the same way as the
magnetic field direction, while $\vec{E}$ and $\vec{B}$ remain transverse and perpendicular.
Therefore, the radial component
of the unmodified expression for the Poynting vector when the source dipole $\vec{p}$ is
linearly polarized is
\begin{equation}
\vec{S}^{(0)}\!\cdot\hat{r}=\frac{1}{2}\Re\left\{\vec{E}\times\vec{B}^{*}\right\}\!\cdot\hat{r}
=\frac{1}{2}\Re\left\{B_{\theta}B_{\theta}^{*}+B_{\phi}B_{\phi}^{*}\right\}\!,
\label{eq-S0}
\end{equation}
which---as is evident from (\ref{eq-B-theta-linear}) and (\ref{eq-B-phi-linear})---is unchanged
from the value taken in conventional, Lorentz-invariant electrodynamics.

However, things may become significantly trickier when the dipole is oscillating elliptically.
In (\ref{eq-B-theta-linear-2}--\ref{eq-B-phi-linear}), the magnetic field amplitudes
[with $e^{i\omega(r-t)}$ factored out] reduced to real expressions, but this will not be
true in the presence of a nonzero $\alpha$. Instead of $p_{2}$,
$e^{i\alpha}p_{2}=p_{2}(\cos\alpha+i\sin\alpha)$ always appears. Reading off,
by analogy with (\ref{eq-B-theta-linear-2}--\ref{eq-B-phi-linear}),
\begin{eqnarray}
B_{\theta} & \propto & \left[(p_{1}\cos\phi+p_{2}\cos\alpha\sin\phi)^{2}\cos^{2}\theta
+(p_{1}\sin\phi-p_{2}\cos\alpha\cos\phi)^{2}\right]^{1/2} \nonumber\\
& & \times\cos\left\{\left(k_{0}r-\vec{k}\cdot\vec{r}\right)-\tan^{-1}\left[
\frac{(p_{1}\cos\phi+p_{2}\cos\alpha\sin\phi)\cos\theta}{p_{1}\sin\phi-p_{2}\cos\alpha\cos\phi}\right]
\right\} \nonumber\\
& & -ip_{2}\sin\alpha\left[\cos\phi\cos\left(k_{0}r-\vec{k}\cdot\vec{r}\right)-\cos\theta
\sin\phi\sin\left(k_{0}r-\vec{k}\cdot\vec{r}\right)\right]
\label{eq-B-theta-ellipse} \\
B_{\phi} & \propto & \left[(p_{1}\cos\phi+p_{2}\cos\alpha\sin\phi)^{2}\cos^{2}\theta
+(p_{1}\sin\phi-p_{2}\cos\alpha\cos\phi)^{2}\right]^{1/2} \nonumber\\
& & \times\sin\left\{\left(k_{0}r-\vec{k}\cdot\vec{r}\right)-\tan^{-1}\left[
\frac{(p_{1}\cos\phi+p_{2}\cos\alpha\sin\phi)\cos\theta}{p_{1}\sin\phi-p_{2}\cos\alpha\cos\phi}\right]
\right\} \nonumber\\
& & -ip_{2}\sin\alpha\left[\cos\phi\sin\left(k_{0}r-\vec{k}\cdot\vec{r}\right)+
\cos\theta\sin\phi\cos\left(k_{0}r-\vec{k}\cdot\vec{r}\right)\right]\!,
\label{eq-B-phi-ellipse}
\end{eqnarray}
the real parts of these expressions are still straightforward; however,
one more set of trigonometric identities are necessary to simplify the imaginary parts.
The amplitudes of the imaginary parts of (\ref{eq-B-theta-ellipse}) and (\ref{eq-B-phi-ellipse})
are
\begin{equation}
p_{2}\sin\alpha\left(\cos^{2}\phi+\cos^{2}\theta\sin^{2}\phi\right)^{1/2}
=\left[\vec{p}_{I}\!^{2}-\left(\vec{p}_{I}\cdot\hat{r}\right)^{2}\right]^{1/2}\!.
\end{equation}
With this simplification, we finally have
\begin{eqnarray}
B_{\theta} & = & \frac{\omega^{2}}{4\pi}\frac{e^{i\omega(r-t)}}{r}\left\{
\left[\vec{p}_{R}\!\!^{2}-\left(\vec{p}_{R}\cdot\hat{r}\right)^{2}\right]^{1/2}
\cos\left[\left(k_{0}r-\vec{k}\cdot\vec{r}\right)+\vartheta_{R}\right]\right. \nonumber\\
& & +\left.i\left[\vec{p}_{I}\!^{2}-\left(\vec{p}_{I}\cdot\hat{r}\right)^{2}\right]^{1/2}
\cos\left[\left(k_{0}r-\vec{k}\cdot\vec{r}\right)+\vartheta_{I}\right]\right\}
\label{eq-Bfinal1} \\
B_{\phi} & = & \frac{\omega^{2}}{4\pi}\frac{e^{i\omega(r-t)}}{r}\left\{
\left[\vec{p}_{R}\!\!^{2}-\left(\vec{p}_{R}\cdot\hat{r}\right)^{2}\right]^{1/2}
\sin\left[\left(k_{0}r-\vec{k}\cdot\vec{r}\right)+\vartheta_{R}\right]\right. \nonumber\\
& & +\left.i\left[\vec{p}_{I}\!^{2}-\left(\vec{p}_{I}\cdot\hat{r}\right)^{2}\right]^{1/2}
\sin\left[\left(k_{0}r-\vec{k}\cdot\vec{r}\right)+\vartheta_{I}\right]\right\}\!.
\label{eq-Bfinal2}
\end{eqnarray}
The phase $\vartheta_{R}$ was previously given, and the phase for the imaginary part
$\vartheta_{I}$ may be determined analogously. Since these expressions separate the dipole
moment $\vec{p}$ only into its real and imaginary parts, and not on the specific planar
form (\ref{eq-p-planar}), they must actually hold for arbitrary $\vec{p}$ (although
with a more general $\vec{p}$
the specific formulas for the phases $\vartheta_{R}$ and $\vartheta_{I}$ would need modification).

From (\ref{eq-Bfinal1}) and (\ref{eq-Bfinal2}),
it is evident that the radial component of the unmodified Poynting vector
(\ref{eq-S0}) is unaffected by $k^{\mu}$ even when the oscillating dipole is generating
elliptically polarized radiation. This is actually not unexpected, for dimensional reasons. $\vec{S}$
has units of energy per unit area per unit time, or $($momentum$)^{4}$. The field strengths $\vec{E}$
and $\vec{B}$ are each proportional to $\omega^{2}$, so there is no way for a product of the
$E$ and $B$ fields we have found to give a quantity with the right dimensions that also depends linearly
on $k^{\mu}$. However, there are still other terms through which the energy and momentum outflows
might be modified.

\section{Further Possibilities for Energy-Momentum Flow}

\label{sec-further}

These additional possibilities come in two types. One of them arises from the fact that,
as already noted, $\vec{S}^{(0)}$ is not the correct expression for the energy flux density or the
momentum density. There are additional terms appearing in (\ref{eq-S}) and (\ref{eq-P}) involving
the potentials $A_{0}$ and $\vec{A}$. Since the potentials are normally on linearly proportional
to $\omega$, and because the novel terms [with the forms
$\vec{\mathcal{P}}^{(1)}=-\vec{k}\left(\vec{A}\cdot\vec{B}\right)$ and
$\vec{S}^{(1)}=-k_{0}A_{0}\vec{B}+k_{0}\vec{A}\times\vec{E}$] also depend
linearly on $k^{\mu}$, we expect that they may give rise to dimensionally correct corrections to the
energy and momentum flow that are proportional to $k\omega^{3}$.

The other way in which additional contributions to the energy and momentum transport might arise would
be if the circular polarization vectors (\ref{eq-epsilon-pm}) were insufficiently accurate. When
$k^{\mu}$ is purely timelike, the $\hat{\epsilon}_{\pm}$ are the exact polarization vectors for
propagating plane wave modes. However, if $\vec{k}\neq0$, these forms are only approximate, and the
corrections can become substantial when $k^{\mu}$ is spacelike, for propagation along directions $\hat{r}$
for which $k^{\mu}\cdot\hat{r}_{\mu}=k_{0}-\vec{k}\cdot\hat{r}$ is small. Possibly compensating for
this, however, is the fact that these are exactly the directions for which the splitting in the dispersion
relation (\ref{eq-Q-pm}) is also small.

One potentially tricky fact about the modified polarization states is that the basis vectors for the
normal modes of propagation do not need to be the same for the $\vec{E}$ and $\vec{B}$ fields. This is
evident, for example, from the divergence equations for plane waves in vacuum. Gauss's law is modified,
\begin{equation}
i\vec{Q}\cdot\vec{E}=2\vec{k}\cdot\vec{B},
\end{equation}
but $i\vec{Q}\cdot\vec{B}=0$ is not. So the polarization basis vectors for $\vec{E}$, but not for
$\vec{B}$, may acquire longitudinal components. Moreover,
the other homogeneous Maxwell's equation (Faraday's law),
\begin{equation}
i\vec{Q}\times\vec{E}=i\omega\vec{B},
\label{eq-Q-cross-E}
\end{equation}still provides the relationship between the
polarization bases for the two fields.

In fact, the exact polarization vectors for arbitrary $k^{\mu}$ are known.
The transverse polarization vectors for the magnetic field
are~\cite{ref-lehnert2,ref-kaufhold,ref-silva1}
\begin{equation}
\hat{\epsilon}_{\pm}^{(B)}\propto \left(\omega^{2}-Q_{\pm}^{2}\right)\hat{\theta}
-2i\left(k_{0}Q_{\pm}-\omega\vec{k}\cdot\hat{r}\right)\hat{\phi}.
\label{eq-epsilon-B}
\end{equation}
However, in order to find the $k$-linear corrections to these polarization vectors, it
is necessary to use a more precise expression for the relationship between $\omega$ and
$Q_{\pm}$. This is discussed in section~\ref{sec-mod-pol}.

\subsection{Explicit Modifications to $\Theta^{\mu\nu}$}

Note that, in general, these two types of pathways for finding contributions to the net energy and momentum outflow
could potentially come into play simultaneously.
However, there is a straightforward power counting argument that the modified polarization structure
cannot, at leading order, play a role in physical contributions from $\vec{S}^{(1)}$ and
$\vec{\mathcal{P}}^{(1)}$.
When expanding all quantities in powers of $k^{\mu}$ and neglecting
all modifications beyond linear order, the explicit presence of a $k^{\mu}$ component in a
formula such as $k_{0}\vec{A}\times\vec{E}$ would mean that we would only need to use the
conventional expressions for $\vec{A}$ and $\vec{E}$, as derived in the Lorentz-invariant
Maxwell theory.
In any case, from the fact that the propagating $\vec{B}$ remains exactly transverse even in the
Chern-Simons theory, we can actually conclude that the $-k_{0}A_{0}\vec{B}$ term can never contribute to
a net energy outflow away from the dipole, since the dot product of this term with $\hat{r}$ is
vanishing.

Nonetheless, it is worth adding a few words about the peculiar character of the scalar potential
in the kind of analysis that we are undertaking. The standard forms for the (seemingly) propagating
part of the scalar and vector potentials in the Lorenz gauge (which is
frequently considered the most convenient for radiation problems) are
\begin{eqnarray}
A_{0} & = & -i\frac{\omega}{4\pi}\left(\hat{r}\cdot\vec{p}\,\right)\frac{e^{i\omega(r-t)}}{r} \\
\vec{A} & = & -i\frac{\omega}{4\pi}\,\vec{p}\,\,\frac{e^{i\omega(r-t)}}{r}.
\end{eqnarray}
For $\vec{A}$, it appears to be straightforward to separate this expression into separate
right- and left-circular polarization modes, which can each then be modified to account for the
nonstandard energy-momentum relation in the Chern-Simons theory. However, this methodology does
not appear to be applicable to $A_{0}$, precisely because it is a scalar quantity with no reference
to polarization directions. Moreover, $A_{0}$ evidently only depends on the radial component of the
$\vec{p}$, whereas in the standard theory the $\hat{r}\cdot\vec{p}$ component of the
dipole moment is precisely the part which does not affect the $\vec{E}$ and $\vec{B}$ radiation fields
in the direction $\hat{r}$. Clearly, an $A_{0}$ with this form cannot contribute to any physically
observable characteristics of the emitted radiation; in fact, the role that $A_{0}$ actually plays
is simply to cancel other equally unphysical contributions to $\vec{E}$ that come from the
longitudinal component of $\vec{A}$. The reason that there is no straightforward separation of
$A_{0}$ into two pieces, associated with the two physical polarization modes, is that $A_{0}$
is actually only associated with the unphysical longitudinal mode---which, quite naturally, does
not even really have a well-defined dispersion relation.
Moreover, had we chosen a transverse gauge with $\vec{\nabla}\cdot\vec{A}=0$,
the far-field, wavelike part of $A_{0}$ would have been vanishing and the issue of trying to
disentangle to right- and left-handed modes in the scalar field would never have arisen.

This, in turn, suggests another interesting possibility. We have already
observed that the radial component of the first term in
$\vec{S}^{(1)}=-k_{0}A_{0}\vec{B}+k_{0}\vec{A}\times\vec{E}$
is necessarily zero, because of the structure
of $\vec{B}$; this is true regardless of what form $A_{0}$ takes and thus whatever
gauge conditions have been imposed. It is then tempting to wonder whether it might be possible,
with a judicious choice of gauge, to make the radial component of the second term also vanish!
However, we shall set this question aside for now, in favor of a direct evaluation of
$\vec{A}\times\vec{E}$ in the Lorenz gauge.

The separation of $\vec{A}$ into its two circularly-polarized components follows straightforwardly
along the same lines as the separation of the magnetic field $\vec{B}$. The analogue of (\ref{eq-B-sep})
is
\begin{equation}
\vec{A}=-i\frac{\omega}{4\pi}\frac{e^{i\omega(r-t)}}{r}
\sum_{\pm}\left(\hat{\epsilon}_{\pm}^{*}\!\cdot\vec{p}\,\right)
e^{\pm i\left(k_{0}r-\vec{k}\cdot\vec{r}\right)}\,\hat{\epsilon}_{\pm}+\vec{A}_{L}.
\label{eq-A-sep}
\end{equation}
Although the two vectors $\hat{\epsilon}_{\pm}$ do not form a complete basis for three-dimensional
space, they do span the two-dimensional space of transverse polarizations. The (longitudinal)
remainder term $\vec{A}_{L}$ is unphysical, with its contributions to the electric field $\vec{E}$
being canceled by those coming from $A_{0}$. In the transverse gauge, $\vec{A}_{L}=0$, and we shall
adopt this simplifying convention henceforth. The full decomposition may be carried out,
but it is simpler to notice that if $A_{0}=0$ (which is the case in the far field if $\vec{A}_{L}=0$
also), then we simply have $\vec{A}=\vec{A}_{T}=-\frac{i}{\omega}\vec{E}$.

Consequently, the outward component of the explicit modified part of the formula for $\vec{S}$ is
\begin{equation}
\vec{S}^{(1)}\!\cdot\hat{r}=\frac{1}{2}k_{0}\Re\left\{-\frac{i}{\omega}\vec{E}\times\vec{E}^{*}\right\}
\!\cdot\hat{r}
=-\frac{k_{0}}{2\omega}\Re\left\{i\left(E_{\theta}E_{\phi}^{*}-E_{\phi}E_{\theta}^{*}\right)\right\}\!.
\label{eq-S1}
\end{equation}
If the dipole is oscillating linearly, so that $\vec{E}$ has a form of a real-valued vector
field times a spherical wave phase factor $e^{i\omega(r-t)}$, then (\ref{eq-S1}) is manifestly
zero. However, for a complex dipole, (\ref{eq-Bfinal1}) and (\ref{eq-Bfinal2}) give
\begin{equation}
\vec{S}^{(1)}\!\cdot\hat{r}=\frac{k_{0}\omega^{3}}{16\pi^{2}r^{2}}
\left[\vec{p}_{R}\!\!^{2}-\left(\vec{p}_{R}\cdot\hat{r}\right)^{2}\right]^{1/2}
\left[\vec{p}_{I}\!^{2}-\left(\vec{p}_{I}\cdot\hat{r}\right)^{2}\right]^{1/2}
\sin(\vartheta_{R}-\vartheta_{I}).
\end{equation}
This looks like the signature of a new effect; however, this modification to the energy flow is
actually illusory. Since $\vec{S}$ and $\mathcal{E}$ are not gauge invariant, the only physically
meaningful measure of energy outflow is the total integral of $\vec{S}\cdot\hat{r}$ over the
sphere at spatial infinity. While $\vec{S}^{(1)}\!\cdot\hat{r}$ may be nonzero in certain directions,
it is odd as a function of angles, so that when integrated over the whole sphere, the result
is necessarily vanishing. This can be seen from the previous expression for $\vartheta_{R}$,
\begin{equation}
\vartheta_{R}=-\tan^{-1}\left[\frac{(p_{1}\cos\phi+p_{2}\cos\alpha\sin\phi)\cos\theta}
{p_{1}\sin\phi-p_{2}\cos\alpha\cos\phi}\right],
\end{equation}
and the similar formula for $\vartheta_{I}$. Comparing the values of $\vartheta_{R}$ in
antipodal directions $(\theta,\phi)$ and $(\theta',\phi')=(\pi-\theta,\phi+\pi)$,
the trigonometric functions $\cos\theta'=-\cos\theta$, $\cos\phi'=-\cos\phi$, and
$\sin\phi'=-\sin\phi$ all change signs, so that $\vartheta_{R}'=-\vartheta_{R}$.
Qualitatively, we might expect to that the radial component of $\vec{S}$ should depend on
the quantity $\left(\vec{p}_{R}\times\vec{p}_{I}\right)\cdot\hat{r}$, which is a pseudoscalar
and so has opposite signs in the $\hat{r}$- and $(-\hat{r})$-directions. This is related to the
fact that the net ellipticity, seen over the entire $4\pi$ range of solid angles, is expected to
be zero, since any right-elliptically-polarized waves emitted along $\hat{r}$ will be
counterbalanced by left-elliptically-polarized emission in the antipodal direction.

For any modified contributions to the momentum
outflow, it turns out that there is a very similar argument.
Notice that in the explicitly $k^{\mu}$-dependent term in (\ref{eq-Theta}), the second index
is simply that of $k^{\mu}$ itself, so the spatial densities $\Theta^{0\nu}$ all have modifications of
the form $-k^{\nu}\vec{A}\cdot\vec{B}$. The corresponding integrated quantities are the components of
the electromagnetic energy-momentum four-vector,
\begin{equation}
P^{\mu}=-k^{\mu}\int d^{3}x\,\vec{A}\cdot\vec{B}=-k^{\mu}H,
\end{equation}
proportional to the total magnetic helicity $H$~\cite{ref-moffat-h,ref-berger-field}.
The relationships between the novel terms in the Poynting vector $\vec{S}$ and the Maxwell stress
tensor is
\begin{equation}
\stackrel{\leftrightarrow}{T}{}={}\stackrel{\leftrightarrow}{T}\!{}^{(0)}+\vec{k}\left(-A_{0}\vec{B}
+\vec{A}\times\vec{E}\right){}={}\stackrel{\leftrightarrow}{T}\!{}^{(0)}+
\frac{\vec{k}}{k_{0}}\vec{S}^{(1)}.
\end{equation}
Since the net momentum loss rate of the radiation is the integral of
$\stackrel{\leftrightarrow}{T}\cdot\,\hat{r}$ over the sphere at infinity, the vanishing of the
integral of (\ref{eq-S1}) over all directions dictates that the integral of the
analogous term for the momentum, $\stackrel{\leftrightarrow}{T}\!{}^{(1)}\cdot\hat{r}$
must also give zero.

\subsection{Modified Polarization Structure}

\label{sec-mod-pol}

This leaves the only possible channel for modifications to the
energy or momentum emission at $\mathcal{O}(k\omega^{3})$
to be through the $k^{\mu}$-modified polarization vectors (\ref{eq-epsilon-B}). Along with
(\ref{eq-epsilon-B}) for $\vec{B}$, there are also the modified polarization vectors for the
electric field~\cite{ref-lehnert2,ref-kaufhold,ref-silva1},
\begin{equation}
\hat{\epsilon}_{\pm}^{(E)}\propto 2\left(k_{0}Q_{\pm}-\omega\vec{k}\cdot\hat{r}\right)\hat{\theta}
-i \left(\omega^{2}-Q_{\pm}^{2}\right)\hat{\phi}
+\left[\hat{\epsilon}_{\pm}^{(E)}\cdot\hat{r}\right]\hat{r}.
\label{eq-epsilon-E}
\end{equation}
The radial term is nonzero, but its effects can be neglected. There are several reasons for this. Firstly,
the radial component is smaller than the other two by a factor of $\mathcal{O}\left(k\omega^{-1}\right)$,
so it only affects the normalization of $\hat{\epsilon}_{\pm}^{(E)}$
[noting that (\ref{eq-epsilon-B}) and (\ref{eq-epsilon-E}) are, as yet, not normalized]
at linear order in $k^{\mu}$. This radial term will also not affect the projection of the standard
$\vec{E}$ (which is transverse, apart from nonpropagating terms that fall off rapidly with distance)
onto $\hat{\epsilon}_{\pm}^{(E)}$. Finally, no radial $\vec{E}$ field can contribute to
$\vec{S}^{(0)}\!\cdot\hat{r}$.

Therefore, to the order of interest, (\ref{eq-Q-cross-E}) still reduces to
$\vec{E}=\vec{B}\times\hat{r}$, and so the expression (\ref{eq-S0}) for the outgoing Poynting
vector is still valid.
However, the previous expressions for $B_{\theta}$ and $B_{\phi}$---derived using (\ref{eq-epsilon-pm})
instead of (\ref{eq-epsilon-B})---are not. Instead, we must apply
\begin{equation}
\vec{B}=\frac{\omega^{2}}{4\pi}\frac{e^{i\omega(r-t)}}{r}
\sum_{\pm}\left[\hat{\epsilon}_{\pm}^{(B)*}\!\cdot\left(\hat{r}\times\vec{p}\,\right)\right]
e^{\pm i\left(k_{0}r-\vec{k}\cdot\vec{r}\right)}\,\hat{\epsilon}_{\pm}^{(B)}.
\end{equation}

Evaluating the components of $\hat{\epsilon}_{\pm}^{(B)}$ directly from (\ref{eq-epsilon-B})
is actually trickier than it looks. Expanding the two components to leading order in $k^{\mu}$
and then normalizing, the $k^{\mu}$-dependence actually cancels out, leaving just
(\ref{eq-epsilon-pm})! Instead, the most straightforward way to evaluate $\hat{\epsilon}_{\pm}^{(B)}$
is to notice that the $\hat{\theta}$- and $\hat{\phi}$-components of (\ref{eq-epsilon-B}) also appear
in the exact dispersion relation (\ref{eq-Q-exact}). Since the ratio of the components is known exactly,
it immediately follows that
\begin{eqnarray}
\hat{\epsilon}_{\pm}^{(B)} & = & \frac{1}{\sqrt{2}}\left[
\left(1-\frac{4\left|\vec{k}\times\hat{r}\right|^{2}}{\omega^{2}-Q_{\pm}^{2}}\right)^{-1/4}\hat{\theta}
\pm i\left(1-\frac{4\left|\vec{k}\times\hat{r}\right|^{2}}{\omega^{2}-Q_{\pm}^{2}}\right)^{1/4}\hat{\phi}
\right]
\label{eq-epsilon-B-exact} \\
& = & \frac{1}{\sqrt{2}}\left[\left(1\mp\frac{1}{2}
\frac{\left|\vec{k}\times\hat{r}\right|^{2}}{k_{0}-\vec{k}\cdot\hat{r}}\right)\hat{\theta}
\pm i\left(1\pm\frac{1}{2}
\frac{\left|\vec{k}\times\hat{r}\right|^{2}}{k_{0}-\vec{k}\cdot\hat{r}}\right)\hat{\phi}\right]\!.
\label{eq-epsilon-B-final}
\end{eqnarray}
The approximation made in the last expression (\ref{eq-epsilon-B-final}) is clearly dicey
in the small ranges of
angles for which $k^{\mu}\cdot\hat{r}_{\mu}$ is very small---$\mathcal{O}(k^{2})$, instead of
merely $\mathcal{O}(k)$. (And obviously, these angular ranges only exist for spacelike $k^{\mu}$.)
However, this should actually not pose a problem, because the birefringence itself vanishes as
$k^{\mu}\cdot\hat{r}_{\mu}\rightarrow0$. Note that the denominators in (\ref{eq-epsilon-B-exact})
are precisely $\omega^{2}-Q^{2}$, so that when these quantities are $\mathcal{O}(k^{2})$, the
right-left difference in phase speeds is also $\mathcal{O}(k^{2})$, meaning the propagation is
conventional at leading order in $k^{\mu}$.

The relative simplicity of the common factors in (\ref{eq-epsilon-B-final}) means that
the polarization vectors take the elliptical forms
\begin{equation}
\hat{\epsilon}_{\pm}^{(B)}=\hat{\epsilon}_{\pm}\mp\left(\frac{1}{2}
\frac{\left|\vec{k}\times\hat{r}\right|^{2}}{k_{0}-\vec{k}\cdot\hat{r}}\right)\hat{\epsilon}_{\mp}
\equiv\hat{\epsilon}_{\pm}\mp\mathcal{T}\hat{\epsilon}_{\mp}.
\label{eq-epsilon-B-mixed}
\end{equation}
With the angular dependence in (\ref{eq-epsilon-B-mixed}), there does not appear to be any prospect for
cancelations between the energy outflow in antipodal directions, since the term in parentheses
(which we have denoted $\mathcal{T}$) does not
simply change sign under $\vec{r}\rightarrow-\vec{r}$---unless, clearly, if $k_{0}=0$. In fact, if
$k^{\mu}$ is spacelike, then there is an observer frame in which $k_{0}$ is indeed vanishing, and this
is precisely the frame in which the stability of the theory is manifest, since the energy density
$\mathcal{E}=\mathcal{E}^{(0)}$ is unmodified, meaning that the total energy is bounded below.

Moreover, since the $\mathcal{O}(k\omega^{3})$ contributions to $\vec{S}$ that come from the
$k^{\mu}$-dependent terms in (\ref{eq-epsilon-B-mixed}) come entirely from the $\vec{S}^{(0)}$ part of
(\ref{eq-S}), these contributions are gauge invariant at the level of $\vec{S}$ itself, rather than
only in integrated form. That suggests that it may be possible to identify the angular distribution of the
emitted radiation, not merely the total rate of power emission. However, the gauge invariance of
$\vec{S}^{(0)}$ does not, on its own, guarantee that it actually has a physical interpretation, since
$\vec{S}^{(0)}$ is not actually
the spatial part of a conserved energy-density current without the
inclusion of the explicitly Lorentz-violating terms in $\Theta^{\mu\nu}$.
%

The triple products needed for calculating $\vec{B}$ with the modified $\hat{\epsilon}_{\pm}^{(B)}$
polarization vectors are again given by (\ref{eq-tri-prod}). The calculation proceeds along the same
lines as in section~\ref{sec-fields}, and the result is the same---no change to the power emitted.
However, there still remains the possibility that the radiation fields may carry away a net momentum,
which can be calculated by integrating $\stackrel{\leftrightarrow}{T}\!{}^{(0)}\cdot\hat{r}$ over
the sphere at spatial infinity. The dot product of the standard stress tensor
$\stackrel{\leftrightarrow}{T}\!{}^{(0)}$ with the radial unit vector is
\begin{equation}
\stackrel{\leftrightarrow}{T}\!{}^{(0)}\cdot\hat{r}=
\vec{E}\left(\vec{E}\cdot\hat{r}\right)+\vec{B}\left(\vec{B}\cdot\hat{r}\right)-\mathcal{E}^{(0)}\hat{r}.
\label{eq-T-dot-r}
\end{equation}
Of the three terms on the right-hand side of (\ref{eq-T-dot-r}), only the first can be associated with a
$\mathcal{O}(\mathcal{T})$ net momentum outflow.

The explicit expression for the longitudinal part of the modified polarization vector (\ref{eq-epsilon-E})
is fairly awkward. However, to leading order in $k^{\mu}$ and the far-field approximation, the radial
component of the electric field may be found simply from the modified Gauss's law (\ref{eq-Gauss}),
which reduces to
\begin{equation}
i\omega E_{r}=2\vec{k}\cdot\vec{B}.
\label{eq-Gauss-reduced}
\end{equation}
Note that this means that the leading contribution to the radial field $E_{r}$ simply depends linearly
on $k^{\mu}$, rather than via the more elaborate nonperturbative quantity $\mathcal{T}$, and this
actually presages the fact that this term too will have a vanishing net contribution to the total
energy-momentum outflow. In fact, there is already a clear issue with the
$\vec{E}E_{r}$ term in $\stackrel{\leftrightarrow}{T}\!{}^{(0)}\cdot\hat{r}$.
Because of the vacuum birefringence, the direction of $\vec{E}$ corkscrews around, varying
with distance. This $r$-dependent behavior means that $\stackrel{\leftrightarrow}{T}\!{}^{(0)}\cdot\hat{r}$
cannot have a gauge-independent interpretation, describing the emission of momentum in different
directions in a fashion than can be verified experimentally. Evaluating
$\stackrel{\leftrightarrow}{T}\!{}^{(0)}\cdot\hat{r}$ at different large $r$ values along a single ray
will give vector expressions pointing different directions. At this stage, however,
this does not necessarily rule
out having a well-defined, nonvanishing, $k^{\mu}$-dependent modification to the
total momentum outflow rate, found by integrating
$\stackrel{\leftrightarrow}{T}\!{}^{(0)}\cdot\hat{r}$ over a sphere at large $r$, if the angular integration
conspires to make the directional variability seen along different rays cancel out, producing an
integrated quantity that does not have such an unphysical dependence on $r$.

However, this turns out not to happen, and instead the momentum outflow rate,
$\stackrel{\leftrightarrow}{T}\!{}^{(0)}\cdot\hat{r}$ integrated over a sphere at $r\rightarrow\infty$,
simply vanishes. This may be seen explicitly by decomposing the expression into Cartesian components---and
the result is actually the most straightforward in the $\vec{k}$-direction, which is precisely
the direction in which we would expect a net momentum transfer to be most likely. (This is the direction in
which a stationary charge emits vacuum Cerenkov radiation, for example.)
In this direction, the key contribution comes from
\begin{eqnarray}
\left(\vec{E}\cdot\vec{k}\right)\!\left(\vec{E}\cdot\hat{r}\right) & = & 
\left[\left(\vec{B}\times\hat{r}\right)\cdot\vec{k}\right]\left(-\frac{2i}{\omega}\vec{B}\cdot\vec{k}\right) \\
& = & -\frac{1}{\omega}\Re\left\{i\left[
k_{\theta}^{2}B_{\phi}B_{\theta}^{*}-
k_{\phi}^{2}B_{\theta}B_{\phi}^{*}
+k_{\theta}k_{\phi}\left(B_{\phi}B_{\phi}^{*}-B_{\theta}B_{\theta}^{*}\right)
\right]\right\}\!.
\end{eqnarray}
The $k_{\theta}k_{\phi}$ term is purely imaginary and so vanishes when the real part is taken. On
the other hand,
the $k_{\theta}^{2}$ and $k_{\phi}^{2}$ terms vanish when integrated over all
angles, just as happened with the expression in (\ref{eq-S1}).

\section{Conclusions and Outlook}

\label{sec-concl}

The net result of our calculations is therefore that the Larmor power emission, which (along with vanishing
net momentum outflow) characterizes the radiation from a classical oscillating dipole, is unchanged
in the Lorentz-violating Chern-Simons theory at leading order in the size of the Lorentz violation
coefficients.
This is not actually extremely unexpected. Indeed, one might be tempted to argue that the vanishing
results ought to follow from the discrete symmetries of the Chern-Simons term alone.
The operator parameterized by $k_{0}$ is odd under parity, while those parameterized by $\vec{k}$ are
odd under time reversal. However, this is probably too facile, since our calculations went beyond
a perturbative power-series expansion in the components of $k^{\mu}$. We considered the
$k^{\mu}r_{\mu}\gg 1$ regime and thus effectively resummed certain terms to all orders in
$k^{\mu}$. This allowed for the inclusion of the key phenomenon of polarization rotation, which we
found---in the discussion of $\stackrel{\leftrightarrow}{T}\!{}^{(0)}$---could actually
lead to useful insights about the extent to which various quantities could be assigned gauge-independent
physical interpretations. Equally interesting was the appearance of the fundamentally nonperturbative
quantity $\mathcal{T}$ in the transverse components of the modified polarization vectors.
[It is the case, however, that whenever a key quantity to be integrated over all angles can be
expressed using a form like (\ref{eq-Gauss-reduced})---with the $k^{\mu}$-dependent right-hand side
taking the form of the first term in an uncomplicated-looking power-series expansion---the
axial vector character of $k^{\mu}$ does guarantee that the integral must vanish.]

The peculiar form of $\mathcal{T}$ offered a possible mechanism for how symmetry-driven angular
cancellations could be evaded---although that mechanism did not actually come into action at the orders
we were considering. All the energetic quantities that involved $\mathcal{T}$ specifically turned out to
have their $\mathcal{T}$-dependent behavior cancel out along each individual emission direction; in spite
of the nonperturbative modifications to the polarization structure, the Poynting vector maintained
the standard form~(\ref{eq-S0}).

Nevertheless, the structure of $\mathcal{T}$ does provide some guidance for understanding how
the radiation structure will be modified at higher orders in the Lorentz violation, and there is no question
that there will be modifications at the next order in $k^{\mu}$; this was already demonstrated with
the calculation of the vacuum Cerenkov emission in the theory with a spacelike $k^{\mu}$. For an
oscillating dipole composed of point charges separated by a characteristic size $d\sim v/\omega$, the
$\mathcal{O}(q^{2}k^{2})$ factor in the Cerenkov emission rate for the individual moving charges is
$\mathcal{O}(p^{2}k^{2}\omega^{2})$---leaving out the
Cerenkov radiation's dependence on the charge velocity $v$, which is
complicated and fundamentally nonperturbative; even a stationary charge may spontaneously radiate
unless $k_{0}=0$.
Thus, the expected $\mathcal{O}(p^{2}k^{2}\omega^{2})$ emission rate behavior found in the Cerenkov
process agrees with what we would expect to find from extending the formalism in this paper to the
next order in $k^{\mu}$; and it may
actually be interesting to connect the formalism we have used to the Cerenkov
radiation calculations by carrying out that extension. Since radiation damping is fairly easily described
in the standard oscillating dipole system, such calculations could provide a way of finding a
generalized radiative friction term applicable in the Lorentz-violating theory.

Moreover, all these questions about radiation are also naturally connected to questions about the behavior
of photonic quanta in the Chern-Simons theory.
The quantization of the theory would lead to further changes in how energy-mommentum is
transported to infinity, although we appear to be limited in our ability to quantize the Chern-Simons
theory, due to the energy instability created by the time component of $k^{\mu}$. The potential
divergence of the energy density can be seen directly in the formula (\ref{eq-E}) for $\mathcal{E}$.
The instability can also be seen by evaluating the group velocity of the negative-frequency
modes~\cite{ref-adam}. To deal with this obstruction,
it appears that we must bound the energy density from below,
so that the quantum states of the theory can be built up in the usual way as excitations atop the vacuum.
In fact, a spacelike CPT-odd axial vector $k^{\mu}$ picks out a preferred frame in which the energetics are
well behaved; the theory may be quantized in the $k_{0}=0$ reference frame. The existence of this
special frame, where the stability of the energy density is manifest from the form of $\mathcal{E}$,
affects the structure of vacuum Cerenkov radiation in the spacelike theory. A charged particle moving
with a velocity $\vec{v}$ which exceeds the phase speed of light will emit radiation,
and the radiation applies a back-reaction force on the charge, which tends to bring the charge to rest
in precisely the frame where $k^{\mu}$ has no temporal component. This connection between the condition
for the energy to become stable and the dynamics of individual charges was a very important motivation for
the current work. It seems that there should also be a connection between the radiative reaction force
on a dipole that is emitting radiation by oscillating, but the order of calculations in this paper has
not been sufficient to capture this phenomenon.

Another obvious direction along which the calculations in this paper could be generalized would be to
look at radiation from higher-$\ell$ multipoles. However, our expectation based on what we have found
for the electric dipole case is that there would again be no leading-order $k^{\mu}$-dependent effects.
Nevertheless, it may be an interesting mathematical physics exercise to generalize the methods we have
utilized in this paper to account for radiation in all possible electric and magnetic multipole modes.

More generally, it is interesting to ask how the elaborate details of radiation theory will change
in the presence of Lorentz-violating preferred backgrounds like $k^{\mu}$. Many useful techniques
have been developed for making quantitative calculations of tricky quantities in Maxwellian electrodynamics
and for understanding their qualitative characteristics. To what extent these methods continue
to be useful---and what adjustments are needed to keep them so---in a theory with exotic modifications
to the charged particle and photon sectors may provide insights both into the specific theories
being considered and the general structure of classical (or quantum) radiation processes.


\end{document}